\documentstyle[12pt]{article}
\topmargin - 1.5cm
\title{Covariant technique of derivative expansion of one-loop
effective action}

\author{N.G. Pletnev\thanks{e-mail: pletnev@math.nsc.ru} and 
A.T. Banin}

\date{{\it
Institute of Mathematics, Novosibirsk, \\
Prosp. Koptiug 4, 630090, Russia}\\
\vspace{1 truecm}
}
\begin{document}
\language=1

 %Inserted by TeXtelmExtel

\begin{titlepage}
\maketitle
\vspace{1 truecm}

 %Inserted by TeXtelmExtel

\begin{abstract}
A simple systematic method for calculating derivative expansions of the one-loop
effec\-tive action is presented. This method is based on using symbols
of operators and well known deformation quantization theory. To demonstrate
its advantages we present several examples of application for scalar theory,
Yang-Mills theory, and scalar electrodynamics. The superspace formulation of
the method is considered for K\"ahlerian and non-K\"ahlerian quantum
corrections for Wess-Zumino and for Heisenberg-Euler lagrangians in super
QED models.
\end{abstract}

 %Inserted by TeXtelmExtel

\thispagestyle{empty}
\end{titlepage}

 %Inserted by TeXtelmExtel

\newcommand{\be}{\begin{equation}}
\newcommand{\ee}{\end{equation}}
\newcommand{\bea}{\begin{eqnarray}}
\newcommand{\eea}{\end{eqnarray}}

 %Inserted by TeXtelmExtel

\section{Introduction}

 %Inserted by TeXtelmExtel

The low energy effective action (EA) (see ref. \cite{Dw}) contains all
predictions of quantum field theory and is a central object of research
in physical situations, when we are interested in phenomena at an
energy scale which is smaller then some cutoff $\Lambda$. Then fundamental
heavy degrees of freedom of the underlying theory  appear in loops
only as virtual states and an integration over both these mass states
and all massless excitations above the scale we are interested in leads, generally
speaking, to nonlocal quantities.

 %Inserted by TeXtelmExtel

Unfortunately, the straightforward calculations even for one-loop EA, determined
by the spectrum of the operator $H = {\delta^2 S \over \delta \phi \delta \phi} $
as a functional of external fields, face this essential problem. Such a problem
can be precisely solved only for some very specific simple
configurations of background fields, when eigenvalues of $H$ can be
found precisely. 
Therefore, the problem of
development of the manifestly covariant methods for calculating the (non-local)
EA as a series of local terms depending on background field derivatives has attracted
much attention. The leading term, named the effective potential, is the most investigated term
in the derivative expansion. It is a useful object for the determination of
vacuum structure of the full theory \cite{It}.

 %Inserted by TeXtelmExtel

The most known method for calculating the derivative expansion EA (DEEA) 
is the so-called Schwinger - DeWitt
asymptotic expansion \cite{Bar, Ball, Schw}. All interesting quantities,
such as EA, Green function, energy-momentum tensor, currents, and
anomalies are expressed in this approach in terms of
asymptotic coefficient heat kernel decomposition, so-called
Hadamard - Minakshi\-sundaram - DeWitt - Seeley (HMDS)
coefficients.  Various effective covariant methods for
calculating HMDS coefficients has been developed by many
authors. Schwinger - DeWitt decomposition gives the good
description of vacuum polarization effects of mass fields on a
background of weak background fields.  However, the description of
such physical phenomena as Hawking radiation or the anomalous magnetic
moment of the electron involves consideration of the nonlocal
structure of the effective action. Among various methods for
investigation of nonlocal effective dynamics, such as direct
summation of the terms with higher derivatives \cite{Avr} and
integration of anomalies, the most preferable is the covariant
method of the theory of perturbations \cite{Vil}.

 %Inserted by TeXtelmExtel

On the other hand, it has been known for a long time, that one-loop EA can be,
at least formally, rewritten as, first quantized path integral for
a fictive particle, which correctly describes the behavior of spin and color
degrees of freedom in external fields \cite{Fr}. This representation and
its modifications are used for calculation of derivative
expansion in QFT, for research of complicated Feynman diagrams
and for application of stationary phase method \cite{Bern}.
Unfortunately, the application of the quantum-mechanical path integral
in curved phase space meets the difficulties of its correct
definition. This is related to the time-slicing
procedure\cite{Van}, because it is not covariant itself.
Moreover, in order to get a sense of the path integral, it is
necessary to add some designations in the way of
finite-dimentional approximation. The ambiguities arising from
such procedure have the same source as the quantization procedure
\cite{Ber}.

 %Inserted by TeXtelmExtel

The powerful nonrenormalization theorems in the supersymmetrical theories
\cite{West} do not prohibit the quantum corrections for superpotential.
So far perturbative calculations determine the effective K\"ahlerian
potential. When the supersymmetry is unbroken, this potential determines both
the effective potential of the theory and the kinetic
terms. The problem of calculating the K\"ahlerian potential was
developed by many authors both on a component level \cite{Vec},
and with the use of the supergraph technique \cite{Gri}.
The generalization of the operator Schwinger-DeWitt representation for
an appropriate heat kernel is also developed (see, for example, ref. \cite{Buch}).

 %Inserted by TeXtelmExtel

The principe of manifest covariance is crucial for effective theories
constructing. It means that any physical theory, which possesses some symmetry,
 must be formulated in such a form where all symmetries are manifest both
at classical and quantum levels. The main advantages of the background
field method consist in the fact that it allows us to formulate 
supersymmetrical and gauge invariant theory of perturbations manifestly(see ref.
\cite{Gat} and reference therein).

 %Inserted by TeXtelmExtel

A lot of interest in perturbative calculations one-loop EA for $N=2,4$
SYM theories has been attracted recently. It was induced by exact
Seiberg-Witten results in $N=2$ supersymmetrical gauge
theories without and with material hypermultiplets \cite{Wit},
where the K\"ahlerian potential and mass of stable states are
predetermined by holomorphity and duality of the prepotential in the
space of quantum modules of the theory.

 %Inserted by TeXtelmExtel

 %Inserted by TeXtelmExtel

An undertaken test \cite{Rocek, Pick} of the forms of non-abelian
supersymmetrical EA by direct calculations indicates
the presence of one-loop holomorphic functions $ {\cal F} $ and
real function ${\cal H}(W,\bar W)$, which are incompatible with special
geometry and consequently with $N=2$ supersymmetry. Therefore a problem of
contributions of higher dimensions and their influence on Bogomol'nyi -Prasad-Summerfield
(BPS) formula of
mass remains important. One of the main obstacles in the investigation of
the EA in $N=2,4$ SYM models in conventional superspace is the presence
of infinitely reducible structure. The formulation of the theory in
harmonic superspace in terms of unconstrained superfields \cite{Gik}
has not quantization problems. Recently, the first examples of quantum
calculations with manifest $N=2$ supersymmetry have been given within
the context of harmonic superspace formalism \cite{Ovr}.

 %Inserted by TeXtelmExtel

There is an unsolved problem of how to break $N=2$ supersymmetry.
The $N=2$ supersymmetry can be broken spontaneously or softly if we
want to save its useful properties.
The soft breaking \cite{Alv} is a very practical approach to
analyzing possible phenomenological applications
of exact solutions. But it has a limited predictive power because of a
plenty free parameters.
Therefore, finding nonsupersymmetrical vacuum solutions for the scalar
potential induced by quantum effects in the hypermultiplet sector of $N=2$
gauge theories is an important problem.

 %Inserted by TeXtelmExtel

However, the above mentioned problems, despite active attention to them recently, 
do not still go beyond an
approximation of constant background fields. In this paper we try
to develop a scheme \cite{Ch} for calculations for one-loop DEEA,
equally suitable for models, which can contain internal symmetry,
and gauge or other background fields or superfields. In ref. \cite{Gai} the
authors have offered this computing scheme for gauge theories, but
really they did not go further then extraction of divergences. We want to
ratify this method as very effective for some problems. It should
be noted that in the refs. \cite{Ch, Gai} the derivative expansion
method was presented as a collection of separate useful expressions
and identities. At the same time the direct connection between them
and the problem of deformation quantization  can be easily seen \cite{Bay}.

 %Inserted by TeXtelmExtel

We use the definition of a star (or Moyal) product \cite{WWGM} to give
a phase space definition of the operator trace. This allows us
to get a convenient derivative expansion for the heat kernel.
The star product approach to quantization is particularly adapted
to such problems. First, its structure allows us to deal with the
expansion in $\hbar$ in a simple way. Secondly, it is the only known
general quantization scheme which allows the quantization of any
symplectic manifold including those where a choice of the polarization
is impossible. Extensive lists of the literature on this subject
can be found in ref. \cite{Stern}.

 %Inserted by TeXtelmExtel

Here we present a covariant method which consists of a sequential
application of the symbol operator technique for formal trace calculation
of the evolution operator. In practice this leads to
a normal coordinate expansion of all quantities contained in the heat
kernel and using the finite translation property of momentum integrals.
This property is also used in other approaches \cite{Nov} to calculate
quantum corrections using a modified propagator, which has all gauge
invariant combinations of background fields and their derivatives
already. It should be noted that this procedure does
not affect the space-time relation of background fields.
The proposed technique allows us to produce a derivative expansion for the
effective action on the background of exact solutions for the Heisenberg
equation.

 %Inserted by TeXtelmExtel

Obtaining that or other specific results has demonstrated the character of
basic elements of the method. We concentrate on advances in
other calculation schemes, examples of the scalar theory with self-action
in flat space \cite{Ch, Fli}, calculation colorless QCD correlators \cite{Nov},
 and simple derivation of the chiral anomaly. We shall consider the problem
of derivative expansion in scalar electrodynamics \cite{Gus}, which is
laying outside the frameworks of   theory of perturbations.
A calculation by a manifestly supersymmetrical way of the first famous correction
to K\"ahlerian potential in Wess-Zumino model \cite{Yar, Pick} and
super\-genera\-lization Schwinger EA in super QED \cite{Shiz} will
be presented also.

 %Inserted by TeXtelmExtel

The plan of the paper is as follows. We begin with a brief consideration
of the offered method. Then we present several examples to demonstrate
its scope for the mentioned problems.
The paper ends with a short summary.

 %Inserted by TeXtelmExtel

 %Inserted by TeXtelmExtel

\section{The method}  % 2
The starting expression for the calculation the one-loop EA, obtained
by integrating over quantum and (or) heavy fields in functional
integral is \cite{Dw}
$\Gamma_{(1)} = -1/2 {\rm Tr} \ln \hat{H} $,
where operator $\hat{H}$ is the second functional derivative of the
action, i.e., the inverse propagator in the presence of background fields.
To give sense of this formal expression we use the known technique of symbols
of operators \cite{Ber}. In this approach the quantum expectation 
value of the operator $\hat{A}$ is 
\be\label{trace}
{\rm Tr}(\hat{A}) = \int_{X} d\mu(\gamma) A(\gamma),
\ee
where $X$ is the phase space and $A(\gamma)$ is the function on the phase 
space (i.e., the symbol of the corresponding operator $\hat{A}$). 
The symbol caclculus is based on the so-called star product which corresponds 
to the usual product of operators.
In this case the standard notation of
one-loop EA in the form of the heat kernel or in the suitable for
regularization $\zeta$ - function form
\be\label{dzeta}
\zeta_{H}(s) = {1 \over \Gamma(s)}\int_{0}^{\infty}dT T^{s-1}
{\rm Tr}({\rm e}^{-T\hat{H}}),
\ee
provides us a connection with Wigner-Weyl-Moyal formalism
\cite{Anton}, since, due to eq. (\ref{trace}) we can write
\be\label{stardzeta}
\zeta_{H}(s) = {1 \over \Gamma(s)}\int_{0}^{\infty}dT T^{s-1}
\int_{X}d\mu(\gamma){\rm e}^{-T H(\gamma)}_{\star},
\ee
where ${\rm e}^{-T H}_{\star}$ is the star exponential, defined by
$$
{\rm e}_{\star}^{H} = \sum {1 \over N!} H\star H\star \cdots H\star.
$$
This allows us to derive phase-space expressions for the formal trace by
$$
Tr(\hat{A})= \int_{X}d\mu(\gamma)(A + \hbar\tau_1(A)+ \hbar^2\tau_2(A)+\cdots)
$$
in quasiclassical expansion form.

 %Inserted by TeXtelmExtel

In order to introduce some notations we will use in the paper we briefly
review the phase space formulation of ordinary quantum mechanics
(originated by Weyl, Wigner, and Moyal \cite{WWGM} and extensively
studied by Berezin \cite{Ber}).

 %Inserted by TeXtelmExtel

Symplectic manifold $X$ can be viewed as a cotangent fiber bundle
$(X^{2n},X^n, T^{\ast}_{x}X,\omega)$ with the base space $X^{n}$,
fiber $T^{\ast}_{x}X$ and fundamental symplectic two-form $\omega$.
 In local coordinates, we have $\gamma=(p_{1}\ldots p_{n},x^{1}\ldots x^{n})$,
$\gamma \in X^{2n}$, $x \in X^n, p \in T^{\ast}_{x}X$,
$\omega={1 \over 2}\omega_{ij}d\gamma^{i}\wedge d\gamma^{j}$. In Hamilton
mechanics $X^{2n}$ plays the role of phase-space equipped by standard
Poisson brackets $\{f,g\}_{\omega}= \omega^{ij}\partial_{i} f \partial_{j} g$.

 %Inserted by TeXtelmExtel

Let us consider some dynamical system on a flat phase space.
Let some quantization be chosen, i.e., linear mapping $A \leftrightarrow
\hat{A}$ between functions in the phase-space (classical observations)
and operators in the Hilbert space by the following recipe
\be\label{map}
A(\gamma) \rightarrow \hat{A} = \int_{X}d\xi d\eta w(\xi,\eta)
\tilde{A}(\xi,\eta){\rm e}^{{i \over \hbar}(\xi \hat{p}-\eta \hat{q})},
\ee
where $\tilde{A}$ is the inverse Fourier transform $A$;
$(\hat{p},\hat{q})= \hat{\Gamma}$ are operators satisfying the
canonical commutational relation $[\hat{p},\hat{q}]=-i\hbar$,
$w(\xi,\eta)$ is a some weight function, which depends on ordering
rule and $(\xi,\eta)= \gamma$  belongs to dual $X^{2n}$ space.
For practical calculations, it is very convenient to employ a
differential form of the last relation, i.e.,
\be\label{difmap}
\hat{A}=A(-i\partial_{\gamma})e^{i\gamma\omega\hat{\Gamma}}
\mid_{\gamma=0}.
\ee
Of course an operator can be characterized by function other then
phase-space based symbol. A prime example is its integral kernel,
i.e. the Dirac matrix element $\langle x\mid\hat{A}\mid y \rangle$
for which the following formulas are useful. Taking the matrix element
of eq. (\ref{map}) leads to a construction of the kernel starting from
the Weyl symbol (i.e., $w(\xi, \eta) = 1$)
\be\label{kern}
\langle x\mid\hat{A}\mid y \rangle = \int dp {\rm e}^{{ip \over \hbar}(x-y)}
A({x+y \over 2},p).
\ee
One passes in the opposite direction from the kernel to symbol via the
Wigner transform
\be\label{covsym}
A(p,q)=\int dv {\rm e}^{-ipv \over \hbar}\langle q+
{1 \over 2}v\mid\hat{A}\mid q-{1 \over 2}v \rangle =
{\langle q\mid\hat{A}\mid p \rangle \over \langle q\mid p \rangle},
\ee
where $\mid p \rangle$ is the momentum eigenstate.
The obtained asymmetric form $A$ is suitable for calculations.

 %Inserted by TeXtelmExtel

As soon as the mapping $A(\gamma) \leftrightarrow \hat{A}$ is constructed,
the star product appears in  phase space, which copies product
of operators. This construction is essentially nonlocal, which is
characteristics of the quantum uncertainty principle. For this basic
structure there are again both integral and derivative based formulas,
which are useful in varying circumstances
\be\label{starint}
(A\star B)(\gamma) =\int\int d\gamma'd\gamma''A(\gamma +\gamma')
B(\gamma +\gamma''){\rm e}^{{2i \over \hbar}\gamma'\omega\gamma''},
\ee
\be\label{stardif}
(A\star B)(\gamma) = {\rm e}^{{i\hbar\over 2}\partial_{\gamma'}
\omega\partial_{\gamma''}}A(\gamma') B(\gamma'')
\mid_{\gamma=\gamma'=\gamma''}= AB + i\hbar\{A,B\}_{PB} + \ldots.
\ee
The Groenewold formula (\ref{stardif}) is a consequence of
eq. ( \ref{starint}) and provides a small $\hbar$ expansion of
$(A\star B)$.
The fact that it may be evaluated through the translation of function
arguments is the key feature
\be\label{starsh}
(A\star B)(p,q) = A(p-{i\hbar \over 2}\partial_q,q+
{i\hbar \over 2}\partial_p)B(p,q).
\ee
The image of commutator in the WWM formalism is  the Moyal bracket $\{A,B\}_M$,
 which is bilinear, skewed and obeys the Jacobi identity.

 %Inserted by TeXtelmExtel

It can be proved that different choices of star product correspond to
different choices of operator ordering. Furthermore,
there is a $W_\infty$ symmetry linking the various choices of the star product.

 %Inserted by TeXtelmExtel

For a dynamical system on curved phase space the above mentioned constructions
assume natural generalizations \cite{Tyt}.

 %Inserted by TeXtelmExtel

Because the correspondence $A \leftrightarrow \hat{A}$ claims on an
autonomous quantum mechanic statement, there has to be correspondence
between physical results for particular dynamic systems. The quantum
equations of motion are then obtained from the classical picture having
pointwise multiplication and Poisson bracket replaced with their star
analo\-gues. It was proved that for an exactly solvable quantum-mechanical
system, the corresponding star analogue of the evolution operator has a
Fourier-Dirichlet expansion
$$
{\rm e}_{\star}^{TH}=\sum_{\lambda \in I}\mid\lambda\rangle\langle\lambda
\mid {\rm e}^{T\lambda \over i\hbar}
$$
This allows us to localize a functional integral, turning it into a sum over
spectrum of the operator $\hat{H}$ \cite{Bay}.

 %Inserted by TeXtelmExtel

In a dynamical system, which does not have the exact solution for the spectrum,
we have to calculate asymptotic expansion coefficients of the heat kernel.
Our suggestion is that  it is convenient to present the star product as
the argument displacement.

 %Inserted by TeXtelmExtel

Though the operator ordering isn't essential, there are a number of
systems having an inherit polarization. For example, if the
$\hat{H}= \hat{p}^2 +V(\hat{q})$ then the $qp$ ordering is the most
preferable and
\be\label{starexp}
{\rm e}_{\star}^{-TH}={\rm e}^{-T(p^2+V(q-i\hbar\partial_p))}.
\ee
This is the simplest case where $(p,q)\in X$ is a symplectic vector space.
In the following sections we will demonstrate the treatment of such
expressions.

 %Inserted by TeXtelmExtel

In more complicated cases such as a particle in external gravitational
and YM fields, which are connections on a principal
bundle over the configuration space $Q$, the theme of strict deformation
quantization was discussed in a number of works \cite{Land}. First,
the gauge invariant definition of Wigner function was studied by
Stratonovich \cite{Strat}. The specific character of such system consists
in the fact that the phase-space of the particle is a Marsden-Weinstein
reduction of $T^{\ast}G$, hence this space can also be considered
as reduced phase-space of a particular type of constrained dynamical
system. Then the quantization corresponds to assigning quantum operators
to be generators of an irreducible unitary representation of the group $G$.
However, there is more then one such representation of the group and
many different inequivalent quantum systems arise from the study of
the same configuration space. Physically, this means that without a
connection we can't separate the particle's external momentum from
its own internal 'position' and 'momentum' which is associated with
the motion on the coadjoint orbit. Using the connection $\nabla$  on
$Q$ we had constructed a star product of standard ordered type
$\star_s$, which is the natural generalization of the standard
ordered product in flat $X$ \cite{Bord}. A surprisingly simple analogue
of the operator
$$
N={\rm e}^{{\hbar \over 2i}{\partial^2 \over \partial_{p_k} \partial_{q^k}}}
$$
 for any
$T^{\ast}Q$ takes the form
$$
N={\rm e}^{{\hbar \over 2i}\Delta}.
$$
Here the second order differential operator $\Delta$ is as follows
$$
\Delta = {\partial^2 \over \partial q^{i} \partial p_{i}} +
\Gamma^{i}_{ik}(q){\partial \over \partial p_{k}} +
p_{k}\Gamma^{k}_{ij}(q){\partial^2 \over \partial p_{i} \partial p_{j}} +
A_k(q) {\partial \over \partial p_{k}},
$$
where $\Gamma^{i}_{jk}$ is the Cristoffel symbol and $A$ is one-form
on $Q$ such that $dA$ equals to the strength tensor.

 %Inserted by TeXtelmExtel

The operator $N$ is globally defined and induces the equivalence
transformation, which yields a more physical star product of Weyl
type having the complex conjugation as an involutive antilinear
antiautomorfism
$$
f\star_W g = N^{-1}((Nf)\star_s(Ng)).
$$
This equivalence is again the natural generalization of the flat case.

 %Inserted by TeXtelmExtel

The facts mentioned above prescribe the following gauge invariant way to
determinate the connection Weyl type symbol (related to $\nabla$ ordering,
because they are not commutative)
\bea\label{Nabla}
\nabla^p_\mu={\rm e}^{i\partial_{p} \cdot\nabla}
( ip_\mu + \nabla_\mu) {\rm e}^{-i\partial_{p} \cdot\nabla} = ip_{\mu}
+ \int^1_0 d \tau\cdot i\tau\partial^\nu_p F_{\nu\mu} (x + i\tau\partial_p) =
\\ \nonumber
=ip_\mu + {i\over 2} \partial^\nu_p F_{\nu\mu} -
{1\over3} \partial^{\lambda\nu}_p
F_{\nu\mu,\lambda} - {i\over 8} \partial^{\sigma\lambda\nu}_p
F_{\nu\mu,\lambda\sigma} + \ldots
\eea

 %Inserted by TeXtelmExtel

The action of the operator $U={\rm e}^{i\partial_{p} \cdot\nabla}$
corresponds to a canonical transformation, which leads to the normal
coordinate expansion. Here a role of tangent vector, along which
implements parallel transport, plays ${\partial \over \partial
p}$, in the $p$ - representation of the normal coordinates.
To find (\ref{Nabla}), we used the commutation relation
$[\nabla_\mu,\nabla_\nu] = F_{\mu\nu}(q)$.

 %Inserted by TeXtelmExtel

For development of the offered technique in superspace where the choice
of gauge condition isn't obvious, we notice that we have obtained
a representation of the vector potential in the Fock-Schwinger gauge
$$
A_\mu(q)=q^\nu\sum^\infty_{n=0}{1\over n + 2}{1\over n!}
q^{\alpha_1}... q^{\alpha_n} F_{\nu\mu,\alpha_1 ...\alpha_n},
$$
 without explicit solving the gauge condition $q^\mu A_\mu=0.$
The potential term is presented by the expression
\be\label{Pot}
V_p={\rm e}^{i\partial_{p} \cdot\nabla} V(q) {\rm e}^{-i\partial_{p}
\cdot\nabla} = V(q + i\partial_p)
\ee
in the normal coordinate expansion form.  Now we get a representation
of the main object for calculations in the form ${\rm Tr} \ln(-\Box_p + V_p).$

 %Inserted by TeXtelmExtel

The main result of the technique of symbols is that already on the first stage
of calculations we have found initial expression for $H(p,q)$ containing
only gauge covariant quantities. The problem of obtaining $\Gamma_{(1)} $,
thus, consists in calculating of the evolution operator of some
quantum-mechanical problem with the hamiltonian $H =-\Box_p + V_p$.
We shall calculate the result of star-product directly, order by order in $T$.
It means that we will implement $p, \partial_p$ ordering until all terms
having derivatives acting on nothing (vacuum) will disappear.
This is quite a simple procedure.  Moreover, the sensible separation $H$
on an exactly soluble hamiltonian $H_0$ and a perturbation $V$ allows us to
construct expansions on the background of $H_0$ eigenstate.

 %Inserted by TeXtelmExtel

 %Inserted by TeXtelmExtel

 %Inserted by TeXtelmExtel

\section{Derivative expansion EA on a background of scalar potential} % 3
As the first example we shall consider a massive scalar field theory with
the lagrangian
$$
{\cal L} = {1 \over 2}\partial_{\mu} \phi \partial^{\mu} \phi -
m^2\phi^2 - U(\phi)
$$
and the problem of the inverse mass decomposition EA for comparison of the offered method
with results string-inspired technique \cite{Fli} and other computing
schemes \cite{Ch, Ball}. Typical problems in which interaction through
derivatives plays an important role are connected with stabilization of
soliton solutions in the Skirm model, QCD and in the Higgs sector of the standard models.

 %Inserted by TeXtelmExtel

It is convenient to use a proper time representation for the
trace of the logarithm of the operator
$\hat{H}=-\Box + V(x), V = m^2 + U''(\phi)$. According to the method we get
the initial representation for one-loop EA
\be
\Gamma_{(1)} =-{1\over 2} \int_{0}^\infty {dT\over T} 
\int d^4 x K(T),
\ee
where
$$
K(T)=
\int {d^4 p\over (2\pi)^4} {\rm e}^{-T (p^2 + V (x + i\partial p))}
$$
is the heat kernel.
The expression for the effective action contains divergences and
imposes renormalization. For the given representation
the $\zeta$ - regularization is intrinsic, which has the advantage of
automatically preserving a large class of the classical symmetries.

 %Inserted by TeXtelmExtel

We leave in the decomposition (\ref{Pot}) fourth order terms in derivatives.
\be\label{Exp}
V_p=V + i\partial^\mu_p V_\mu- {1\over 2} V_{\mu\nu} \partial^{\mu\nu}_p
- {i\over 3!} V_{\mu\nu\lambda} \partial^{\mu\nu\lambda}_p + {1\over 4!}
V_{\mu\nu\lambda\tau} \partial^{\mu\nu\lambda\tau}_p.
\ee
The further problem is calculation of a trace of the
evolution operator for a fictive particle in the potential (\ref{Pot}).
Using the known operator identity
\be\label{Oper}
{\rm e}^{-T (p^2+ V_p)} ={\rm e}^{-Tp^2} \exp\int^T_0 d\tau\,
{\rm e}^{+ \tau p^2} (-V_p){\rm e}^{-\tau p^2},
\ee
the kinetic term can be separated. As a result the argument
of $V_p$ is shifted as $V_p(x + i\partial_p-2i\tau p)$. We shall
consider eq. (\ref{Pot}) as a perturbation and we shall decompose the 
$T$ exponent up to derivatives of fourth order
\be\label{Pexp}
P_{T} \exp\left(\int_{0}^{T} ds V_{p}(s) \right) = \sum_{n=0}^{\infty}
\int_{0}^{T} ds_{1} \int^{s_{1}}_{0} ds_{2}
\ldots\int_{0}^{s_{n-1}} ds_{n} V_{p} (s_{1}) V_{p} (s_{2}) \ldots
V_{p} (s_{n}).
\ee

 %Inserted by TeXtelmExtel

Expressions such as ${\rm e}^ {+ \tau p^2} \partial_ {p} {\rm e}^ {-\tau p^2}$
are replaced with the solutions of the Heisenberg equations, i.e.
$\partial_p(\tau) =\partial_{p} -2\tau p$.  The most complicate procedure
is the disentangling of the star-product. The partial  simplification can
be reached after commutation $\partial_p$ to the left and using properties
\be\label{prop}
\int {d^4 p\over (2\pi)^4} \partial^\mu_p (...) =0.
\ee
All other calculations reduced to trivial integrations
\be
\int {d^4 p\over (2\pi)^4} e^{-Tp^2} \{1, p_\mu p_\nu, p_
{\mu_1}... p_ {\mu_4} \} = {1\over (4\pi T)^2} \{1, {1\over 2T}
\delta_{\mu\nu}, {1\over 4 T^2} \delta_{\mu_1\mu_2\mu_3\mu_4} \}.
\ee
For example,
\bea\label{means}
 \langle \partial^{\nu\mu}_p \rangle \equiv \int {d^ {4} p \over (2\pi)^4}
{\rm e}^{-Tp^2}\int_{0}^{T} ds\partial^{\nu}_p(s) \partial^{\mu}_p(s) =
- {1 \over 3} T^2\delta_{\mu\nu},\\ \nonumber
\langle \partial^{\nu\tau\lambda\rho}_p \rangle = {2 \over 15}
T^3\delta^{\nu\lambda\tau\rho},
\quad \langle \partial^{\mu_{1} \ldots\mu_{6}}_p \rangle = - {2 \over 35}
T^3\delta^{\mu_{1} \ldots\mu_{6}},\\ \nonumber
\langle\langle \partial^{\nu}_p \partial^{\tau}_p \rangle\rangle \equiv
\int {d^ {4} p \over (2\pi)^4} {\rm e}^{-Tp^2}
\int_{0}^{T} ds\partial^{\nu}_p(s) \int_{0}^{s}
ds'\partial^{\tau}_p(s') = - {1 \over 12} T^3\delta^{\nu\tau},
\nonumber
\eea
where $\delta^{\mu_{1} \ldots\mu_{2k}} $ is a completely
symmetrical tensor, consisting of $ (2k-1)!!$ terms
composed from Kronecker symbol products.  After rearranging 
the results by extracting full derivatives, we obtain the
known result \cite{Fli}
\bea
K(T)={1\over (4\pi T)^2} {\rm e}^{-VT}\times
\phantom{vvvvvvvvv}\\ \nonumber 
\times\left(1- {1\over 12}
T^3 V_\mu V_\mu + {1\over 5!} T^4 V_{\mu\nu} V_{\mu\nu} -
{T^5\over 3\cdot 4!} V_{\mu\nu} V_\mu V_\nu + {T^6\over 12\cdot
4!} V_\mu^2 V_\nu^2 \right).
\eea
Further integration over
proper time gives gamma functions in any order of $T$. They have
poles for some terms in DEEA, which correspond to known
divergences.

 %Inserted by TeXtelmExtel

Previously, it was mentioned that the local Schwinger-DeWitt
expansion describes the vacuum polarization effect of massive
quantum fields in weak background when all their invariants are
smaller then the corresponding power of the mass parameter.
However it isn't a good approximation for the case of strong
background fields and absolutely meaningless for massless
theories and weak rapidly varying background fields. For
investigation of these cases special methods are needed
\cite{Avr, Vil}. The result has an essentially nonlocal form.
It is interesting to study how some nonlocal formfactors appear in
our approach for this model. Let's consider the second order
$V$ term in the Duhamel expression (\ref{Pexp}).
After simple manipulation we obtain
\be
K(T) = \int_{0}^{T}ds_{1}\int_{0}^{s_{1}}ds_{2}V(x)
e^{(s_{1}-s_{2})(\Box+2ip\nabla)}V(x)
e^{-(s_{1}-s_{2})(\Box+2ip\nabla)}.
\ee
Performing the integration over $p$ we can see that
the result can be rewritten in the form
$$
K(T)= {1 \over (4\pi T)^2} \int_{0}^{T}ds_{1}
\int_{0}^{s_{1}}ds_{2}Ve^{(s_{1}-s_{2})(1- {s_{1}-s_{2} \over T})\Box}V.
$$
From the last expression we obtain
\be
\Gamma_{(1)} \sim -{1 \over 2}\int {dT \over T} e^{-m^2T}
\int d^4 x {1 \over (4\pi T)^2}{T^2 \over 2 }
V\gamma(T \Box)V,
\ee
where the formfactor has the known representation \cite{Avr}
\be
\gamma(T \Box) = \int_{0}^{1}ds e^{{1- s^2 \over 4}T \Box}
\ee

 %Inserted by TeXtelmExtel

 %Inserted by TeXtelmExtel

Using a similar expression for formfactors, one can analyze
their analytical properties, calculate their high energy limits
and their imaginary parts above the threshold, etc.

 %Inserted by TeXtelmExtel

 %Inserted by TeXtelmExtel

\section{Non-abelian gauge fields} % 4
Now we consider the gauge vacuum mean of gluon operators
in the form $\langle F^2\rangle, \langle F^3\rangle$. In the
absence of consistent theory of the QCD vacuum, it was assumed that
vacuum expectation values of local operators, in fact, play the role
of fundamental constants for QCD sum rules.
The necessity to calculate the coefficients in front of these gluon operators
in decomposi\-tion for colorless correlation functions (which is
used in a method of QCD sums rules) stimulated the development of
gauge-invariant methods \cite{Nov, Ball}. Unfortunately, in the
standard Feynman diagrams technique, calculating the diagrams
with emitting gluons from a loop and rearranging vector potentials
in gauge invariant structures, is a rather hopeless problem.
Because the determinant of the Dirac operator is determined
after squaring by the Klein-Gordon operator, we limit ourselves
to the consideration of a scalar loop in the external nonabelian field
with the lagrangian
$$
{\cal L}={1 \over 2}\nabla_{\mu} \phi \nabla^{\mu} \phi + m^2 \phi^2;
\quad \hat{H}=-\nabla^{\mu}\nabla_{\mu} + m^2.
$$
According to the prescription described above, we use a representation
of one-loop EA (\ref{stardzeta}) with
$H(p,\partial_p)=\nabla^{\mu}_{p}\nabla_{p}^{\mu}+m^2$, where
$\nabla^{\mu}_{p}$ is a covariant pseudodifferential operator (\ref{Nabla}).
After extracting a free part from $H$ in the form $H_0=p^2$ (which corresponds to
the approximation where the particle motion between the interaction
is free) and separating exponents similar to those in eq. (\ref{Oper}) we shall
calculate vacuum mean dimensions $(-3)$.
Nonzero contributions from  (\ref{Pexp}) in the procedure
described in the second section will give the
following results in the first and in the
second order of the decomposition of the $T$ exponent, respectively 
$$
- {1\over 4} F_{\nu\mu} F_{\tau\mu}
\langle \partial^{\nu\tau}_p \rangle +
{ 1\over 72} F_{\nu\mu,\tau\lambda} F_{\sigma\mu}
\langle \partial^{\nu\lambda\tau\sigma}_p \rangle,
\quad
{ 1\over 9} F_{\nu\mu,\mu} F_{\tau\alpha,\alpha}
\langle\langle \partial^{\nu\tau}_p \rangle\rangle.
$$
The other terms are either full derivatives or contain as a factor 
from the left or right $p^{\mu}\partial^{\nu}_p F_{\mu\nu}$, 
which leads
to zero contributions. Further, performing trivial calculations
similar to (\ref{means}) and using the Bianchi identities we will
get the known result \cite{Nov, Ball}
\be
K(T)={ {\rm e}^{-Tm^2} \over (4\pi T)^2} \left(1 + {T^2\over 12} F_{\mu\nu}
F_{\mu\nu} + T^3\left({F^3\over 180} - {1\over 60} J^2\right) \right),
\ee
where $F^3=F_{\mu\nu} F_{\nu\alpha} F_{\alpha\mu}$,
$J^2=F_{\mu\alpha,\alpha} F_{\mu\beta,\beta}$. Thus the first
(after  unit) term of the decomposition is related to renormalization
of charge.

 %Inserted by TeXtelmExtel

A less trivial problem is the calculation of the next HMDS
coefficient $b_{3}\sim\left({F \over m} \right)^4$.
The simplification is reached on free equations of motion
$F_{\mu\nu,\mu} =0$. In the first order of the T -exponent
decomposition we obtain the following contribution
$$
- {1 \over 4\cdot 6!} F_{\alpha\mu} F_{\nu\mu\rho\sigma\tau\lambda}
\langle \partial^{\lambda\tau\sigma\rho\nu\alpha}_p \rangle =
{ T^{4} \over 70\cdot 6!} F_{\alpha\mu} F_{\nu\mu\rho\sigma\tau\lambda}
\delta^{\lambda\tau\sigma\rho\nu\alpha},
$$
where only 10 of 15 members are nonvanishing. After some manipulations
with commutating of derivatives, using of Bianchi identities  and
equations of motions we get the contribution
\be
{ 5T^4 \over 7\cdot 2 \cdot 6!}
\left(
[ F_{\alpha\beta}, F_{\mu\beta}]^2 +
{ 1 \over 10}
[ F_{\alpha\mu}, F_{\rho\sigma}]^2
\right).
\ee
The contribution of the second order of the decomposition (\ref{Pexp}) is
\be
{ T^4 \over 2\cdot 6!} \left(\{F_{\alpha\mu},F_{\beta\mu}\}^2 +
5 (F_{\nu\mu} F_{\nu\mu})^2\right)
\ee
The full result for $b_3$
\be
{ T^4 \over 2\cdot 144} \left(
( F_{\nu\mu} F_{\nu\mu})^2 + {1 \over 5}
 \{F_{\alpha\mu}, F_{\beta\mu} \}^2 +
 {1 \over 7} [F_{\alpha\beta}, F_{\mu\beta}]^2 +
 {1 \over 70} [F_{\alpha\mu}, F_{\rho\sigma}]^2
 \right)
\ee
coincides with \cite{Nov}. It should be noted that the huge number of
 terms in the decomposition can be omitted at once,
that essentially reduces body of work and demonst\-rates that
computation of higher power corrections might be considerably simplified.
This is important for the analysis of convergence for the series in
$1/ m^2$.

 %Inserted by TeXtelmExtel

We have considered two well known examples. Less trivial application
of the EA expansion was used in ref. \cite{Mke} for the investigation of
axial anomaly. The problem of genera\-liza\-tion world line path
integral representation \cite{FG, Hok} for amplitu\-des involving
axial vector leads to another interesting application of the derivative
expansion. It is well known that if the spinor fields are coupled
to background fields $A_{\mu}, A_{5\mu}$ and the pseudoscalar one than
the axial current $J^{5}_{\mu}$ has an anomalous divergence. The Dirac
operator, suitably continued to Euclidian space, isn't Hermitian and
the anomaly can be attributed to the phase of the functional determinant.
In \cite{Sal} using the integral representation of the complex power
for the pseudodifferential operator, it was obtained an unambiguous
definition of the determinant for the Dirac operator.
The determinant is shown to be vector gauge invariant and it posseses
the correct axial and scale anomalies.

 %Inserted by TeXtelmExtel

Another popular starting point is the second order description for
the fermionic one-loop effective action
$$
\Gamma_{(1)} \sim {1 \over 2}\{{\rm Tr}\log HH^{\dagger} +
{\rm Tr}(\log H - \log H^{\dagger})\}.
$$
The derivative of the second term in the last expression with
respect to the background field can be written as
$$
{\delta \over \delta A}{\rm Tr}(\log H - \log H^{\dagger}) =
{\rm Tr}
\{({\delta H \over \delta A}H^{\dagger} -
H{\delta H^{\dagger} \over \delta A} ){1 \over HH^{\dagger}}
\},
$$
which allows us to derive an elegant representation with the help of an
auxiliary integration for the imaginary part of the effective action,
i.e., for the phase of the fermion functional determinant.
Recently \cite{Mke}, it was found that  for the special case where
the background consists only with an abelian vector and an axial
vector field there is a much simpler solution to this problem which 
treats both parts of the effective action equally. 
The price which we
have to pay for this property is that the kinetic operator occurs
non-Hermitian. We consider a more complicated example contained
general nonabelian fields $A_{\mu}^{a}, A_{5\mu}^{a}$. It is easily to
establish that
\be (\not \! p + \not \!\! A+\gamma_{5}\not \!\!
A_{5})^2= -(\partial_{\mu}+i{\cal A}_{\mu})^2 +Q,
\ee where
${\cal A}_{\mu}= A_{\mu}-\gamma_{5}\sigma_{\mu\nu}A_{5}^{\nu},
\quad A=A^{a}T^{a},$ and
\be
Q = -{i \over 2}\sigma_{\mu\nu}F_{\mu\nu} +
i\gamma_{5}A_{\mu,\mu}^{5}+2A^{2}_{5} + {1 \over 2}\sigma^{\mu\nu}
[A_{5\mu},A_{5\nu}],
\ee
with
$$
F_{\mu\nu}= \partial_{\mu}A_{\nu} - \partial_{\nu}A_{\mu} +
i[A_{\mu},A_{\nu}].
$$
Using such a trick, the effective action is formally identical with
the effective action for a scalar loop in non-abelian field ${\cal A}$
and potential $Q$ background. A new gauge parameter has values in the
Clifford algebra. Let's apply the method, described above for the calculation
of the quantity
$$
\Gamma_{(1)} = -{1 \over 2}\int{dT \over T}e^{-Tm^2}
\int d^4x{d^4p  \over (2\pi)^4}{\rm e}^{-T(\nabla_{p}^2 + Q)}.
$$
Repeating the above calculation for this case, we get the known results
$$
K(T)={1 \over (4\pi T)^2}{\rm tr}\{1-TQ + T^2 ({1\over 2}Q^2 - {1 \over 12}{\cal F}^2) +
$$
\be
+ T^3(-{1 \over 6}Q^3-{1 \over 12}Q_{\mu}Q_{\mu} +
{1 \over 12}Q {\cal F}^2 - {i \over 180}{\cal F}^3 + {1 \over 60 }J^2) \}
\ee
where

 %Inserted by TeXtelmExtel

$$
{\cal F}_{\mu\nu} = F_{\mu\nu} +\gamma_{5}\sigma_{[\mu\lambda}
A^{\lambda}_{5,\nu]} + i \sigma_{[\mu\lambda}\sigma_{\nu ]\tau}
A^{\lambda}_{5}A^{\tau}_{5},
$$

 %Inserted by TeXtelmExtel

$$
A^{\lambda}_{5,\mu}=\partial_{\mu}A_{5}^{\lambda} + i[A_{\mu},A_{5}^{\lambda}].
$$

 %Inserted by TeXtelmExtel

We shall need to perform the Dirac traces. We see that unlike the
vector case, the axial contribution to the imaginary part has an
additional term proportional to $m^2 A^{2}_{5}$. This term is prohibited
by gauge invariance in the vector case, however this term  may appear in
axial theories with massive fermions since those theories violate the
corresponding gauge invariance.
Logarithmically diveregent terms combine automatically in the gauge
invariant expressions
$$
K(T)\sim T^2{2 \over 3}Tr_{c}\{G^{A}_{\mu\nu}G^{A}_{\mu\nu} +
G^{A_{5}}_{\mu\nu}G^{A_{5}}_{\mu\nu} \},
$$
where $G^{A}_{\mu\nu}= F_{\mu\nu}+i[A_{5\mu},A_{5\nu}]$.
Therefore it is necessary only to introduce the kinetic and mass
counterterms for the axial field in order to render the theory
be finite.  The third HMDS coefficient contains a lot of terms.
Keeping only the contribution which comes from the three point
function $\langle AAA_{5}\rangle$, we get the famous result 
\be
K(T)\sim -T^3 {\rm Tr}_{c}[G^{A}_{\alpha\beta}G^{A*}_{\alpha\beta}A^{5}_{\mu ,\mu} +
{4 \over 3}\{ G^{A}_{\alpha\beta},G^{A*}_{\beta\gamma}\}
A^{5}_{\gamma,\alpha} ]
\ee
It has shown that the effective action induced by a spinor loop can
be rewritten in terms of an auxiliary nonabelian gauge field and a
potential. This allows us to discuss the chiral anomaly from a novel
point of view.

 %Inserted by TeXtelmExtel

 %Inserted by TeXtelmExtel

\section{Derivative expansion of EA  in QED}  % 5
In recent years a lot of problems
related to intensive fields and non-linear processes such as photon
splitting, non-linear Compton effect, and pair production below the two
photon threshold \cite{Ph} were experimentally investigated. So far
the problem of going beyond perturbation theory increased so much,
that the description of quantum processes becomes rather urgent and
gets practical goals. Really, studying the limit of a strong field
we obtain the same information, as from a polarization function
in the small distance limit.

 %Inserted by TeXtelmExtel

Unfortunately, the validity of the famous Schwinger lagrangian \cite{Schw}
calculated almost a half-century ago and the two-loop exact results
\cite{Rit} are limited by the constant field approximation.
The generalization of the Schwinger result on strong varying fields
or fields located in small area is very interesting from the physical
point of view. Recently the authors \cite{Gus} presented the 
next (after Schwinger term) nonpertur\-bative term
$F_{\alpha\beta,\lambda} F_{\sigma\delta,\gamma}
L_1^{\lambda\alpha\beta\gamma\sigma\delta} (F_{\mu\nu}) $ in the expansion
of one-loop EA. Their result was obtained from the representation
of the path integral. Even for electrodynamics it is a rather difficult problem.

 %Inserted by TeXtelmExtel

In this section we would like to demonstrate the capabilities of the
proposed method for the computation of the complete form for the
first nontrivial correction to long wavelength limit of the EA.
We use the representation (\ref{Nabla}) for the pseudodifferential
operator $\nabla_{p}$ and the proper time representations for EA,
induced by a scalar loop.

 %Inserted by TeXtelmExtel

Let's consider this example in more detail. For the calculation of the expansion
on a nonpertur\-bative background it is necessary to split out free and
perturbation terms in the expression $e^{-T (\Pi^2 + \Delta^{(1)})}$, where
\be
\Pi_\mu=p_\mu + {1\over 2} \partial^\nu_p F_{\nu\mu},\quad
[ \Pi_\mu,\Pi_\nu] =-F_{\mu\nu},
\ee

 %Inserted by TeXtelmExtel

\be
\Delta^{(1)} = {i\over 3} F_{(\nu\mu,\tau)} (2 {\Pi_\mu}
\partial^{\tau\nu}_p
+ \delta^\tau_\mu\partial^\nu_p)
- {1\over 8} F_{(\nu\mu,\tau\lambda)}
( 2\Pi_{\mu} \partial^{\lambda\tau\nu}_p
+ \delta^{(\tau}_{\mu}\partial^{\lambda) \nu}_p)
- {1\over 9} F_{(\nu\mu,\tau} F_{\rho\mu,\lambda)}
\partial^{\tau\nu\lambda\rho}_p
\ee

 %Inserted by TeXtelmExtel

Here parentheses means symmetrization with the appropriate weight.
The interested terms in the expansion of $T$ exponent (\ref{Pexp}) for the
heat kernel are
\bea\label{Xp}
K(T)={\rm e}^{-T\Pi^2} \{1 + \int_0^T ds {\rm e}^{s\Pi^2}
[{1\over 8} F_{\nu\mu,\tau\lambda}
( 2\Pi_\mu\partial^{\lambda\tau\nu}_p +...)
+ {1\over 9} F_{\nu\mu,\tau} F_{\rho\mu,\lambda}
\partial^{\tau\nu\lambda\rho}_p] {\rm e}^{-s\Pi^2} \\ \nonumber
- {1\over 9}
F_{\nu\mu,\tau} F_{\alpha\beta,\gamma}
\int^T_0 ds\,\int_0^s ds' {\rm e}^{s\Pi^2}
(2\Pi_\mu\partial^{\tau\nu}_p +\ldots) {\rm e}^{-(s-s') \Pi^2}
( 2 \Pi^{\beta} \partial^{\alpha\gamma}_p +\ldots) {\rm
e}^{s'\Pi^2} \}.
\eea

 %Inserted by TeXtelmExtel

The following step consists of replacing
${\rm e}^{s\Pi^2} {\Pi_\mu} {\rm e}^{-s\Pi^2},
{\rm e}^{s\Pi^2} \partial_p^\mu {\rm e}^{-s\Pi^2} $ in appropriate
solutions $\Pi_\mu (s),\partial_p^\mu (s)$  of equations of motion
$$
\dot{\Pi}(s) = [\Pi^2,\Pi (s)], \dot{\partial_p^\mu}(s) =
[\Pi^2,\partial_p^\mu (s)],
$$
i.e.,
\be
\Pi_\mu (s) =\Pi^\nu P_{\nu\mu} (s), \quad
\partial^\mu_p (s) =\partial^\mu_p + {\Pi^\nu} B_{\nu}^{\mu} (s),
\ee
where $P (s) = ({\rm e}^{-2sF}) $, $B(s)= {1\over F} ({\rm e}^{-2sF} -1).$
After $ \Pi,\partial_p $ ordering it is necessary to take integrals
over $p$. In principle, some methods for solution of this problem
has already been used by Schwinger.
Recently to a similar problem have addressed the paper \cite{Art},
with the reference to heat kernel calculation methods developed
in \cite{Avrd}.

 %Inserted by TeXtelmExtel

To treat the first term in (\ref{Xp}) we notice that
the operator $\Pi^2$ is the hamiltonian of the two Landau
oscillators in momentum representation. The kernel of the
operator $\langle p' \mid {\rm e}^ {-T\Pi^2} \mid p\rangle$ is
well known explicit Meller formula, frequently used for direct
calculation of the index of an operator. It is important, that
this kernel is a well converging expression and
consequently\footnote{Using this property, the authors
\cite{Art} have reproduced the Schwinger result.}
\be\label{prope}
\int {d^4 p\over (2\pi)^4} \partial^a_p ({\rm e}^{-T\Pi^2} \Pi_\mu...) =0.
\ee

 %Inserted by TeXtelmExtel

 %Inserted by TeXtelmExtel

For future convinience we define the moments
$$
K_ {a_1 a_2... a_n} =
\int {d^4 p\over (2\pi)^4} {\rm e}^{-T\Pi^2} \Pi_{a_1}
\Pi_{a_2}\ldots \Pi_{a_n}.
$$
In particular, for $n=2$ we have
$$ 0=\int {d^4 p\over (2\pi)^4} \partial^b_p ({\rm e}^{-T\Pi^2}
\Pi_a) = \delta_{a}^{b} K + K_{ca} B_{c}^{b}.  $$ The expansion
of a matrix $B$ begins with unit, therefore one can be inversed
$B^{-1}$ and 
$$ 
K_{ab} =-KB^{-1}_{ba}.  
$$ Similarly 
$$
K_{a_1... a_4} =K (B^{-1}_{a_2 a_1} B^{-1}_{a_4 a_3} + B^{-1}_{a_3
a_1}
B^{-1}_{a_4 a_2} + B^{-1}_{a_4 a_1} B^{-1}_{a_3 a_2}).
$$
We also need in the relations
$$
K_{a_1 a_2... a_6} =- (B^{-1}_{a_2 a_1} K_{a_3... a_6} + B^{-1}_{a_3 a_1}
K_ {a_2 a_4... a_6 +...}).
$$
The kernel $K(T)$ satisfies the differential equation
$$
{ dK\over dT} =-K_{aa} =KB^{-1}_{aa},
$$
or
\bea
K^{-1} {dK\over dT} = {\rm tr} ({F \over {\rm e}^{-2TF} -1}) = {\rm tr}
({F{\rm e}^{2TF} \over 1-{\rm e}^{2TF}}) =\\ \nonumber
= - {1\over 2} tr((1-{\rm e}^{2TF})^{-1} {d\over dT} (1-{\rm
e}^{2TF})) = - {1\over 2} {\rm tr} {d\over dT} \ln (1-{\rm e}^{2TF})
C, \nonumber
\eea
here $C=2\pi/F$ is a constant of integration,
determined from a known limit $K=1/ (4\pi T)^2$ for small $T$,
when the particle can be considered as free.  With such a choice
we find the standard result
\be\label{Sch}
K(T)= {1\over (4\pi T)^2} [\det {FT\over \sinh (FT)}]^ {1/2}
\ee

 %Inserted by TeXtelmExtel

As a next step, it is necessary to calculate several functions
from matrixes $F_{\mu\nu}$. It is known \cite{Lan} that for any
constant field $F_{\mu\nu}$ there is such a reference frame, where
magnetic and electrical fields are parallel and their values in this
system are relativistic invariants of the field. Or, if they are
perpendicular, it is possible to find such a reference frame, in
which the field is either purely magnetic or purely electrical.
Therefore the canonical form $F$ in this system has a block structure
$$
F_{\mu\nu}=\pmatrix{0&\lambda_1&0&0\cr
-\lambda_1&0&0&0\cr
0&0&0&-\lambda_2\cr
0&0&\lambda_2&0\cr}.
$$
There is a connection between eigenvalues and invariants of the
field
\be\label{ev}
{\cal H}_{\pm}= (\lambda_{1}\pm i\lambda_{2})^2 =
{1 \over 2}(F^2 \mp i F^*F).
\ee
Any degree $F$ can be decomposed over basis of linear combinations of
$F, F^*, F^2$ and $g$. Thus, for the exponent from a matrix $P={\rm
e}^{\alpha F} $ we get
$$
P={\rm e}^{\alpha\lambda_1}A_1+ {\rm
e}^{-\alpha\lambda_1}A_2+ {\rm e}^{i\alpha\lambda_2}A_3+ {\rm
e}^{-i\alpha\lambda_2}A_4,
$$

 %Inserted by TeXtelmExtel

where $A^{(i)}_{\mu\nu} $ is another known basis \cite{Bat}
$$
A^{(1)}={1\over 2(\lambda^2_1+\lambda^2_2)}
(F^2+\lambda^2_2 g+\lambda_1 F-\lambda_2 F^*)={1\over 2}
\pmatrix{1&1&0&0\cr -1&-1&0&0&\cr 0&0&0&0\cr 0&0&0&0\cr}
$$
$$
A^{(2)}={1\over 2(\lambda^2_1+\lambda^2_2)}
(F^2+\lambda^2_2 g-\lambda_1 F+\lambda_2 F^*)={1\over 2}
\pmatrix{1&-1&0&0\cr 1&-1&0&0&\cr 0&0&0&0\cr 0&0&0&0\cr}
$$
$$
A^{(3)}={-1\over 2(\lambda^2_1+\lambda^2_2)}
(F^2-\lambda^2_1 g+i\lambda_2 F +i\lambda_1 F^*)=-{1\over 2}
\pmatrix{0&0&0&0\cr 0&0&0&0&\cr 0&0&1&-i\cr 0&0&i&1\cr}
$$
$$
A^{(4)}={-1\over 2(\lambda^2_1+\lambda^2_2)}
(F^2-\lambda^2_1 g-i\lambda_2 F -i\lambda_1 F^*)=-{1\over 2}
\pmatrix{0&0&0&0\cr 0&0&0&0&\cr 0&0&1&i\cr 0&0&-i&1\cr},
$$
which has the useful projector properties $A^2_{(i)} =A_{(i)}$,
$A_{(i)} A_{(j)} =0$ for $i\not =j$. The transposition operation
translates $A_1\leftrightarrow A_2$ and $A_3\leftrightarrow A_4$.

 %Inserted by TeXtelmExtel

Calculation of matrix functions $B$ and $B^ {-1} $ leads to
remarkably simple results
\be
B=\sum^4_{i=1}A^{(i)}{1\over f_i}({\rm e}^{\alpha f_i}-1), \quad
B^{-1}=\sum^4_{i=1}A^{(i)}f_i({\rm e}^{\alpha f_i}-1)^{-1}.
\ee
It is convenien to use the notations
$$
f_{1,2} =\pm\lambda_1,\quad f_{3,4} =\pm i\lambda_2.
$$
Now we can easily get
$$
{\sinh(\alpha F)\over \alpha F}={\sinh(\alpha\lambda_1)\over \alpha\lambda_1}
\pmatrix{1&0&0&0\cr 0&-1&0&0\cr 0&0&0&0\cr 0&0&0&0\cr}
+{\sin(\alpha\lambda_2)\over \alpha\lambda_2}
\pmatrix{0&0&0&0\cr 0&0&0&0\cr 0&0&-1&0\cr 0&0&0&-1\cr}
$$
and for the kernel we obtain the Schwinger result
\be\label{Schw}
K(T)={1\over(4\pi T)^2}{T\lambda_1\over \sinh(T\lambda_1)}{T\lambda_2\over
\sin(T\lambda_2)}
\ee

 %Inserted by TeXtelmExtel

Then we can implement all necessary substitutions,
$\Pi,\partial_p$ ordering of the operators and integration over
momenta in the other terms of expression (\ref{Xp}). After some
manipulation, all matrix structures $P, B, B^{-1} $ depending on
$s, s', T$ are grouped in several combinations. The main group is
\be
B^T(s')B^{-1}(T)B(s)-B^T(s')=
-\sum_i{2\over f_i}A^{(i)}{\rm e}^{f_i(s'-s)}\sinh(f_i s'\,)
{\sinh(f_i(s-T))\over \sinh(f_i T)}
\ee
which coincides with the Green function, used in the path integral
method  \cite{Gus}
\bea
D(s,s')=\sum {A^{(i)}\over 2f_i}[-(1-{\rm e}^{2f_i(s'-s)})+
+\coth(f_i T)(1+{\rm e}^{2f_i(s'-s)})-\\ \nonumber
-{1\over \sinh(f_i T)}({\rm e}^{f_i(2s'-T)}+{\rm e}^{f_i(-2s+T)}))].
\eea
The other arising combinations of matrix structures are derivatives
of $D$ in $s, s'$, that can be easily seen
\bea
B^T(s')B^{-1}(T)B(s')-B^T(s')=
\sum_{i}-{2\over f_i}A^{(i)}\sinh( f_i s'\,)
{\sinh(f_i(s'-T))\over \sinh(f_iT)}=D(s',s'),  \\ \nonumber
B^T(s')B^{-1}(T)P(s)=
\sum_i A^{(i)}{\rm e}^{f_i(s'-2s+T)}\,{\sinh(f_i s')\over \sinh(f_i T)}
=-{1\over 2}{\partial\over\partial s}D(s,s'),  \\ \nonumber
P^T(s')B^{-1}(T)B(s)-P^T(s')=
\sum_i A^{(i)}{\rm e}^{f_i(2s'-s)} {\sinh(f_i(s-T)) \over \sinh(f_i T)}
=-{1\over 2}{\partial\over\partial s'}D(s,s'),\\ \nonumber
B^T(s')B^{-1}(T)P(s')=
\sum_i A^{(i)}{\rm e}^{-f_i(s'-T)}\,{\sinh(f_i s')\over \sinh(f_i T)}=
-{1\over 2}{\partial\over\partial s}D(s,s')\vert_{s=s'},\\ \nonumber
P^T(s')B^{-1}(T)P(s)=
\sum_i -{A^{(i)}f_{i}{\rm e}^{f_i(2s'-2s+T)}\over 2\sinh(f_i T)}=
\left(-{1\over 2}{\partial\over\partial s'}\right)
\left(-{1\over 2}{\partial\over\partial s}\right)
D(s,s').
\eea

 %Inserted by TeXtelmExtel

In these notations the result for the expression in the braces
(\ref{Xp}) looks as follows
\bea\label{Der}
&1 +& \int_{0}^{T}ds(
{1\over 8}F_{\mu\nu,\tau\mu}D_{\nu\tau}(s,s) +
{1\over 4}F_{\mu\nu,\tau\lambda}
(
\dot{D}_{\nu\mu}D_{\tau\lambda}+
\dot{D}_{\tau\mu}D_{\nu\lambda}+
\dot{D}_{\lambda\mu}D_{\nu\tau})(s,s)  \\[2 mm]\nonumber
&+&{1\over 9}F_{\nu\mu,\tau}F_{\rho\mu,\lambda}
(
D_{\nu\tau}D_{\rho\lambda}+
D_{\rho\tau}D_{\lambda\nu}+
D_{\rho\nu}D_{\lambda\tau}
)(s,s)
)     \\[2 mm]\nonumber
&+&{4 \over 9}F_{\nu\mu,\tau}F_{\alpha\beta,\gamma}
\int_{0}^{T}ds\int_{0}^{s}ds'
(
\dot{D}_{(\tau\mu}(s,s)
(
\dot{D}_{(\gamma\beta}(s',s')D_{\alpha)\nu)}(s,s') + \acute{D}_{\beta\nu)}(s,s')D_{\alpha\gamma}(s',s')
)      \\[2mm]\nonumber
&+& \dot{D}_{(\alpha\mu}(s,s')
(
\acute{D}_{\beta(\tau}(s,s')D_{\gamma)\nu)}(s,s') +
\dot{D}_{\gamma)\beta}(s',s')D_{\nu\tau}(s,s)
) \\[2mm]\nonumber
&+&
\dot{\acute{D}}_{\beta\mu}(s,s')
(
D_{(\gamma\tau}(s,s')D_{\alpha)\nu}(s,s') +
D_{\nu\tau}(s,s)D_{\alpha\gamma}(s',s')
)
)
\eea
The last step is the calculation of a plenty of standard integrals such as
$$
\int D = T^2\sum_{i}A^{(i)}L(f_{i},T), \quad
\int DD = T^3 \sum_{i,j}A^{(i)}\times A^{(j)}
\{L(T f_{i})L(Tf_{j})+
{L(T f_{j})-L(T f_{i}) \over (T f_{i})^2 - (T f_{j})^2}\}.
$$
Because of combersome of the general result, we do not present
it here. Besides it is inconvenient in a particular physical
problem, where it is necessary only some terms.
Let us note only, that functions of proper time $T$ and relativistic
invariants of fields setting in front of every possible contractions
$F_{\mu\nu,\tau\lambda},$
$F_{\mu\nu,\tau} F_{\alpha\beta,\gamma} $ and
with direct products $A^{(i)}, A^{(i)}
\times A^{(j)}, A^{(i)} \times A^{(j)}
\times A^{(k)} $ are combinations of Langevin
functions $L (x) = {x \coth(x) -1 \over x^2} $ and they are
presented in paper \cite{Gus}.

 %Inserted by TeXtelmExtel

Furthermore, it is necessary to implement renormalization through
the subtraction based on common principle, which requires putting
in zero the radiation corrections at the switched off field as in
the original Schwinger paper \cite{Schw}, and replacing all bare
charges and fields through the physical. Therefore it is easier to
return to the initial expressions and to execute all manipulations
with necessary accuracy.

 %Inserted by TeXtelmExtel

When the mass of the scalar particle is greater than all other scales
of the theory, we can limit the expansion by the next terms to unit
\be
K(T)={T^3 \over 15}\left({1 \over 3}F_{\mu\nu,\lambda}F_{\mu\nu,\lambda} +
{1 \over 2}F_{\nu\mu}F_{\nu\mu,\lambda\lambda}\right).
\ee
This result agrees with \cite{Hauk}. Recently, similar methods for
calculation of corrections to the long wavelength limit of EA on
Yang-Mills background fields was used in \cite{Garg}.

 %Inserted by TeXtelmExtel

It is obvious that the expressions (\ref{Der}) for the description of particular
processes in nonconstant background fields are exact in mass of a
charged particle and field strength. The gradient corrections are
very important for the analysis of the effective potential, since
they can reduce energy of the ground state.

 %Inserted by TeXtelmExtel

This detail presentation evidently demonstrates possibilities to
obtain the corrections on background which possess exact solution
of classical problem. Because of a large number of physical set up of
problems in nonconstant background fields, it is useful to have in
an arsenal of tools of their solution a method, which is alternative
to path integral representation.

 %Inserted by TeXtelmExtel

 %Inserted by TeXtelmExtel

 %Inserted by TeXtelmExtel

\section{Quantum corrections in Wess-Zumino model}  % 6
We demonstrate how to apply the proposed technique to
calculation DEEA for the supersymmetrical theories in
the superfield approach.
The doubtless advantage of the offered method is that
this method does not required the determination of many various
Green functions for calculation of functional trace of the
appropriate heat kernel.  To show it, we obtain the known
K\"ahlerian potential of Wess-Zumino model \cite{Rocek, Pick, Yar}
and lowest order nonk\"ahlerian contributions to one-loop
effective potential.

 %Inserted by TeXtelmExtel

The Wess-Zumino theory described by the action
$$
S(\phi,\bar\phi)=\int d^8 z\bar\phi\phi+\int d^6 z
\left({m\phi^2\over 2}+
{g\over 3!}\phi^3 \right) + h.c.
$$
is a good model for test of various supersymmetric methods,
since it has all specific peculiarities of the theories with
chiral fields, and it enters as an inherent ingredient in
many superfield theories \cite{Gat, Bag}.

 %Inserted by TeXtelmExtel

It is known that a problem of a definition of a superfield EA
agreed with symmetry of the theory can be very effectively solved
in the framework of proper time superfield technique \cite{Buch,
Yar}. For functional integration over quantum chiral fields
,which arise after splitting of fields on quantum and background
ones, it is convenient to introduce unlimited superfields
$\phi=\bar D^2\psi$ and $\bar\phi=D^2\bar\psi$. In principle this
introduce a new gauge invariance into the action, but in the absence
of background gauge fields, the ghost associated with this gauge
fixing are decoupled. Another procedure transforming  the path
integral over the chiral superfields into a path integral over
general superfields has been developing in \cite{Buch}.  The
functional integration over $\psi, \bar{\psi}$ leads to a
determination of
the effective action in the form
$ -{1\over 2}Tr\ln(\hat{H}(x,\theta,D))
$
with the kinetic operator for the given model
\be\label{ham}
\hat{H}=\pmatrix{\lambda&\bar D^2\cr D^2&\bar\lambda\cr} \pmatrix{\bar
D^2&0\cr 0& D^2\cr},\quad\mbox{ where }\quad
\lambda=m+g\phi_{(BG)}.
\ee

 %Inserted by TeXtelmExtel

Except a functional trace, the operation $Tr$ means a matrix
trace as usual.  There are many techniques of calculation
the K\"ahlerian potential which is an analogue of the
Coleman-Weinberg potential \cite{Pick, Yar}.

 %Inserted by TeXtelmExtel

We can implement the Fourier transformation in superspace, though
it isn't necessary, since the $\delta$-function
of Grassmanian coordinates is explicitly known, as well as the action
the $D,\bar D$ derivatives on

 %Inserted by TeXtelmExtel

\be\label{Delta}
\delta(z-z')=\int{d^4 p\over(2\pi)^4}d^2 \psi d^2\bar{\psi}
{\rm e}^{i(x-x')\cdot p
+\psi^\alpha(\theta-\theta')_\alpha
+\bar{\psi}^{\dot{\alpha}}(\bar\theta-\bar\theta')_{\dot{\alpha}}}.
\ee

 %Inserted by TeXtelmExtel

We use the superspace agreements from ref. \cite{Gat} and
we will omit the obvious indexes.
Commutating exponents on the left through the differential
operators we find in the coincidence limits the
standard replacements

 %Inserted by TeXtelmExtel

\be
D_{\theta}=\partial_\theta+i/2 \bar\theta\partial\rightarrow
\psi-{1\over 2} p\bar\theta+D_{\theta},
\quad
\bar{D}_{\bar{\theta}}=\partial_{\bar\theta}+i/2 \partial\theta \rightarrow
\bar\psi -{1\over 2} \theta p+\bar{D}_{\bar{\theta}}.
\ee
To obtain the covariant symbols of the operators $D,\bar D$ in
 momentum representation we use identities

 %Inserted by TeXtelmExtel

\bea\label{Gr}
U(D_{\theta}+\psi-{1\over 2}\bar\theta p)U^{-1}=
\psi-{1\over 2}\partial_{\bar\psi} p=D_p,\\ \nonumber
U(\bar{D}_{\bar{\theta}}+\bar\psi-{1\over 2}\theta p)U^{-1}=
\bar\psi-{1\over 2}\partial_\psi p=\bar D_p,
\eea
where parallel translation operator was chosen in the form
\be\label{trans}
U={\rm e}^{i\partial_p \cdot \partial_x}
{\rm e}^{1/2\theta p\partial_{\bar\psi}-1/2 \partial_\psi
p \bar\theta}{\rm e}^{\partial_\psi D_{\theta}+
\partial_{\bar\psi}\bar{D}_{\bar{\theta}}}
\ee
The anticommutator $\{D_p,\bar D_p\} =-p$ and, naturally, all
 useful algebraic relations for $D_p$ have the same form as in
 $D_{\theta} $  algebra. In addition,
we have a transformation for a general superfield
\be\label{Gra} \phi^p=U\phi U^ {-1} =\phi (x +
i\partial_p,\theta + \partial_\psi,\bar\theta +\partial_
{\bar\psi})
\ee
which is the finite degree polynomial in
$\partial_\psi,\partial_{\bar\psi} $ with factors
$D_\theta\ldots D_ {\bar\theta} \phi$.

 %Inserted by TeXtelmExtel

Let us note that other arrangement of exponents in eq. (\ref{trans})
related to the corresponding replacement of the normal coordinates.
For example, the same transformations with
$$
U={\rm e}^{i\partial_p \cdot \partial_x}
{\rm e}^{\partial_\psi D_{\theta}}
{\rm e}^{1/2\theta p\partial_{\bar\psi}-1/2 \partial_\psi
p \bar\theta}
{\rm e}^{\partial_{\bar\psi}\bar{D}_{\bar{\theta}}}
$$
give us
$$
\psi-\partial_{\bar\psi} p=D_p,\quad
\bar\psi=\bar D_p.
$$

 %Inserted by TeXtelmExtel

The steps described above from the operators to the pseudodifferential
operators on the phase superspace are conventional (see ref. \cite{Gai}).
It should be mentioned, that the final result for the trace of the
operator does not depend on selection (\ref{Delta}) which reflects
the chosen ordering scheme. The replacements (\ref{Gr}), (\ref{Gra})
actually correspond to the transition from the operators to their
symbols and can be justified by the arguments similar to those
described in the second section.

 %Inserted by TeXtelmExtel

Limiting ourselves to a problem of calculation of the first
correction to the potential in decomposition over Grassmanian derivatives,
we split the pseudodifferential operator $H$, acting on phase 
superspace, in two parts
\be
H=H_0+\pmatrix{\Lambda\bar D^2_p & 0\cr 0& \bar\Lambda D^2_p\cr},
\ee
where $\Lambda=\partial^\alpha_\psi D_{\theta^\alpha} \lambda$,
$\bar{\Lambda} =\partial^{\dot {\alpha}}_{\bar\psi}
\bar{D}_{\theta^{\dot {\alpha}}} \bar\lambda$ and $H_{0}$ copies the
form (\ref{ham}). In the following steps we will write  $D, \bar{D}$
instead of $D_{p}, \bar{D}_p $. This must not confuse, because $D_{\theta},
\bar{D}_{\bar{\theta}}$ are contained in $\Lambda, \bar{\Lambda}$ only.
Then the K\"ahlerian potential and the correction can be split
$$
{\rm Tr}\ln H_0+{\rm Tr}\ln\left(1+\pmatrix{\Lambda\bar D^2 & 0\cr 0& \bar\Lambda D^2\cr}
H_0^{-1}\right).
$$
For calculation of ${\rm Tr}\ln (H_0) $ we take out and omit the
'free' part of the operator
$$
\pmatrix{0&\bar D^2 D^2\cr D^2\bar D^2 & 0\cr}.
$$
It is clear, that in the expression
\be
{\rm Tr}\ln \left(1+\pmatrix{0&{1\over D^2\bar D^2}\bar\lambda D^2\cr
{1\over \bar D^2 D^2}\lambda \bar D^2 & 0\cr}\right)
\ee
the nonzero contribution will give only even degrees of the logarithm
decomposition. Unfolding the matrix part of the trace we get
$$
- {1\over 2} \ln \left(1- {1\over D^2\bar D^2} \bar\lambda D^2 {1\over \bar D^2
D^2}
\lambda\bar D^2- {1\over \bar D^2 D^2} \lambda\bar D^2 {1\over D^2\bar D^2}
\bar\lambda D^2 \right).
$$
Further, using the decomposition of unit in front of the logarithm
in the form
$$
1 = {\{D^2,\bar D^2\}-D^\alpha\bar{D}^2 D_\alpha \over \Box}
$$
one can convert all spinor derivatives in "boxes"
$\Box=-p^2$. After that we obtain
\be
K^{(1)}=\int{d^4 p\over (2\pi)^4}{1\over 2p^2}\ln(1+{\lambda\bar\lambda\over
p^2}),
\ee
which gives the known result \cite{Yar, Pick} after integration
and renormalization of a wave function by the condition ${
\partial^2 K \over \partial \phi \partial \bar{\phi}}
\vert_{\phi =\phi_0; \bar{\phi}= \bar{\phi}_0} = 1$.

 %Inserted by TeXtelmExtel

For calculation of the next nonvanishing contribution in the EA
expansion, we rewrite $H^{-1}_0$ in the form
$$
\pmatrix{{1\over D^2\bar D^2 }D^2 & 0\cr
0 & {1\over \bar D^2 D^2}\bar D^2\cr}
\pmatrix{-{1\over\Box_+}\bar \lambda &{1\over \Box_+} \bar D^2\cr
{1\over\Box_-}D^2 &-{1\over \Box_-}\lambda\cr},
$$
where
$\Box_{+} =\bar{D}^2 D^2-\lambda\bar{\lambda} $,
$\Box_{-} =D^2\bar{D}^2-\lambda\bar{\lambda} $.
Then
\be
\pmatrix{\Lambda\bar D^2 & 0\cr 0&\bar\Lambda D^2\cr}(H_0^{-1})
=\pmatrix{-\Lambda P_2{1\over \Box_+}\bar\lambda & \Lambda\bar D^2
{1\over\Box_-}\cr\bar\Lambda D^2{1\over\Box_+} & -\bar\Lambda P_1{1\over\Box_-}\lambda\cr},
\ee
where
$P_1= {D^2\bar D^2\over\Box} $, $P_2= {\bar D^2 D^2\over \Box} $
are the projectors in momentum representation.
The first nonvanishing contribution in the decomposition of the
logarithm gives trace of the fourth degree of the matrix (we keep in
mind the properties of integration over $d^2 \psi d^2\bar\psi$).

 %Inserted by TeXtelmExtel

Moreover, among 16 terms the zero contribution automatically comes from 
 terms containing powers more than two of $\Lambda,
 \bar \Lambda$ and also from terms containing $\bar\Lambda D^2$
and $\Lambda\bar D^2$ from the right, because the derivatives
$\partial_{\bar\psi}$ and $\partial_\psi$ contained in
$\Lambda,\bar{\Lambda}$  act on nothing.  
We are left with 
$$
{\lambda\bar\lambda\over 4\Box\Box^4_\lambda}(\Lambda\bar D^2
D^2\bar\Lambda\bar\Lambda\Lambda\bar D^2 D^2+\bar \Lambda D^2 \bar
D^2\Lambda\Lambda\bar\Lambda D^2\bar{D}^2),
$$
$$
\Box_\lambda = \Box - \lambda\bar{\lambda}.
$$
We shall transfer $\partial_{\psi}, \partial_{\bar{\psi}}
$ to the right, using Heisenberg relation $\{ \partial_\psi, D_p\} =1$.
The trivial integration over Grassmanian and usual momenta
gives us immediately the known result for the non K\"ahlerian
terms \cite{Pick, Yar}, leading to quantum deformations of
classical vacuum of the theory
\be
F^{(1)}={1\over 3\cdot
2^7}\,{D^{\alpha}\lambda D_{\alpha}\lambda \bar{ D}^{\dot\alpha}
\bar{\lambda} \bar{D}_{\dot\alpha}\bar{\lambda} \over
\lambda^2\bar{\lambda}^2},
\ee
where the factor $2^{-7}$ is caused by the superagreements. This kind
of one-loop quantum correction is called the effective potential of auxiliary
fields.
Certainly, such quantum correc\-tions are important in $N=1,2$
supersymmetrical models, since they lead potentially  to the
removal of degeneration in classical vacua of the theory.
This method should be quite general and has important
applications for other interesting cases, for example for models
with explicitly broken supersymmetry.

 %Inserted by TeXtelmExtel

 %Inserted by TeXtelmExtel

 %Inserted by TeXtelmExtel

\section {Heisenberg-Euler lagrangian in SQED}  % 7

 %Inserted by TeXtelmExtel

 %Inserted by TeXtelmExtel

In this section we develop manifestly supersymmetrical gauge invariant
strategy of calcu\-lations of one-loop effective action for the most
general renormalizable $N=1$ models including Yang-Mills fields and
chiral supermuliplets
$$
{\cal S}={\rm tr} \int d^6 z W^2 +
\int d^8 z \bar{\Phi}e^{V}\Phi + [\int d^6 z P(\Phi)+ h.c.]
$$
In more detail we consider the one-loop diagrams only with external abelian
superfields and the expansion in terms of spinor covariant
derivatives of superfields $W, \bar{W} $ which can not be reduced to
usual space-time derivatives. This approximation corresponds to
generalization of the Heisenberg-Euler Lagrangian of usual QED. The
background field method in superspace \cite{Buch,Gat} allows us to treat
both vector supermultiplets and matter superfield on the equal 
footing and in an explicitly gauge-invariant way.
However, in contrast to ordinary gauge theories the gauge connections
are not independent objects and are expressed in terms of the prepotentials.

 %Inserted by TeXtelmExtel

The basics of the method in its "quantum-chiral background-vector"
representation are given in ref. \cite{Gat}.
This approach implies that higher loop contributions can be arranged
in such a way that background fields appears in $\nabla_{A}, W_{A}, \Phi$ only.
After expansion of full action, including gauge-fixing and ghost terms,
in powers of quantum fields, the quadratic part determines a matrix
of the kinetic operator acting in the space of all quantum fields.
The physical quantities depend on particular gauge invariant combinations
of the gauge superfields only, such as the field strength and derivatives
thereof.

 %Inserted by TeXtelmExtel

As in the previous section, the replacement of the operators
by their symbols gives $\nabla\to\psi-
{ 1\over 2} p\bar\theta + \nabla$ with manifest dependence on
grassmanian coordinates. To obtain gauge-invariant and
manifestly supercovariant symbols of operators, we use
identities (\ref{Gr}) with replacement of the flat $D$'s by
covariant ones.

 %Inserted by TeXtelmExtel

Using the known notations and conventions from ref. \cite{Gat}, we
find the expansion of the symbols $\nabla^p=U (\psi- {1\over 2}
p\bar\theta + \nabla) U^{-1}$ in superspace "normal"
coordinates \be\label{nabpsi} \nabla^p_\alpha=
\psi_\alpha-{1\over 2}\bar\partial^{\dot\alpha}p_{\alpha\dot\alpha}+
{i\over 4}\bar\partial^{\dot\alpha}
(\partial^{\dot\beta}_\alpha f_{\dot\beta\dot\alpha}+
\partial^{\beta}_{\dot\alpha} f_{\beta\alpha})-
{1\over 3}\partial_\alpha\bar\partial^{\dot\alpha}i\bar W_{\dot\alpha},
\ee
$$
+{1\over 3}\bar\partial^2 iW_\alpha+{1\over 4}\partial_\alpha\bar\partial^2
D'+{3\over 4!}\bar\partial^2\partial^{\beta}if_{\beta\alpha}+...
$$

 %Inserted by TeXtelmExtel

$$
\nabla^p_{\dot\alpha}=
\psi_{\dot\alpha}-{1\over 2}\partial^\alpha p_{\alpha\dot\alpha}+
{i\over 4}\partial^\alpha
(\partial^{\dot\beta}_\alpha f_{\dot\beta\dot\alpha}+
\partial^{\beta}_{\dot\alpha} f_{\beta\alpha})
+{1\over 3}\partial^2 i\bar W_{\dot\alpha}
$$
$$
-{1\over 3}\bar\partial_{\dot\alpha}\partial^\alpha iW_\alpha
-{1\over 4}\bar\partial_{\dot\alpha}\partial^2 D'
+{3\over4!}\partial^2\bar\partial^{\dot\beta}if_{\dot\beta\dot\alpha}
+...
$$

 %Inserted by TeXtelmExtel

We do not specify here obvious indexes $\psi,\bar\psi, p$  in
the $\partial$ representation of normal super\-coor\-di\-nates. The
quantities $f$, $D'$ are the standard notation for superfields
$f_{\alpha\beta} = {1\over 2} \nabla_{(\alpha} W_{\beta)},$
$D'=- {i \over 2} \nabla^\alpha W_\alpha,$
$\nabla^\alpha W_\alpha + \bar\nabla^{\dot\alpha} \bar W_{\dot\alpha} =0.$
Here the dots mean the expansion in $\nabla_{\alpha\dot\alpha} $
derivatives, which we shall omit keeping in mind problems on the
constant background which is independent on space-time
coordinates, but with arbitrary dependence on Grassma\-nian
coordinates. By the construction, the normal coordinate
expansion used gives the connection decomposition in the Wess-Zumino
gauge.

 %Inserted by TeXtelmExtel

Similarly, for a vector derivative we have
\be\label{nabp}
\nabla^p_{\alpha\dot\alpha}=
i p_{\alpha\dot\alpha}+{1\over 2}(\partial^\beta_{\dot\alpha}f_{\alpha\beta}
+\partial^{\dot\beta}_{\alpha} f_{\dot\alpha\dot\beta})+
\partial_\alpha\bar W_{\dot\alpha}
+\bar\partial_{\dot\alpha} W_\alpha+
{1\over 2}(\partial_\alpha\bar\partial^{\dot\beta}
f_{\dot\beta\dot\alpha}
+\bar\partial_{\dot\alpha}\partial^{\beta}f_{\beta\alpha})+i
\partial_\alpha\bar\partial_{\dot\alpha}D'.
\ee
It is not difficult to check up the validity of the identical
correspondence of the algebra of covariant symbols to the algebra of
covariant derivatives
\be
\{\nabla^p_\alpha,\nabla^p_{\dot\alpha}\}=
i\nabla^p_{\alpha\dot\alpha},
\quad
[\nabla^p_{\alpha\dot\alpha},\nabla^p_{\beta\dot\beta}]=
i(C_{\dot\beta\dot
\alpha}f_{\beta\alpha}+C_{\beta\alpha}f_{\dot\beta\dot\alpha}),
\ee

 %Inserted by TeXtelmExtel

$$
[\nabla^p_{\dot\beta},\nabla^p_{\alpha\dot\alpha}]=
C_{\dot\beta\dot\alpha}W^p_\alpha, \quad
[\nabla^p_{\dot{\alpha}},W^{p}_{\alpha}]=0,
$$
where
$W^p_\alpha=UW_\alpha U^{-1} =W_\alpha +  \partial^{\beta}
f_{\beta\alpha}-i\partial_\alpha D'$. This is the verification
that the gauge connection given by (\ref{nabpsi}), (\ref{nabp})
indeed gives rise to the field strength.

 %Inserted by TeXtelmExtel

It is convenient to present $\nabla^p_ {\alpha (\dot\alpha)} $
in a remarkably simple form $$ \nabla^p_\alpha=
\tilde{\psi}_{\alpha}+{i\over 2}\bar\partial^{\dot\alpha}
\nabla^p_{\alpha\dot\alpha}, \quad
\bar{\nabla}^p_{\dot\alpha}=
\tilde{\bar{\psi}}_{\dot{\alpha}}+{i\over 2}\partial^{\alpha}
\nabla^p_{\alpha\dot{\alpha}},
$$
where
$$
\tilde{\psi}_{\alpha}=
\psi_\alpha+{1\over 3!}\partial_\alpha\bar\partial^{\dot\alpha}i\bar
W_{\dot\alpha}-{1\over 3!}\bar\partial^2 iW_\alpha- {1\over
8}\bar\partial^2\partial^\beta if_{\beta\alpha}- {1\over
4}\partial_\alpha\bar\partial^2 D' $$

 %Inserted by TeXtelmExtel

$$
\tilde{\bar{\psi}}=
\bar\psi+{1\over 3!}\bar\partial_{\dot\alpha}\partial^\alpha i
W_\alpha-{1\over 3!}\partial^2 i\bar W_{\dot\alpha}-
{1\over 8}\partial^2 \bar\partial^{\dot\beta}
if_{\dot\beta\dot\alpha}+{1\over 4!}\bar
\partial_{\dot\alpha}\partial^2 D'.
$$

 %Inserted by TeXtelmExtel

So, we have obtained the connection decomposition in normal
supercoordinates, which naturally can be called a
super\-genera\-lization of the Fock-Schwinger gauge. For some
disscus\-sion about this subject see ref. \cite{Ohr}.

 %Inserted by TeXtelmExtel

Let us consider a particular example of calculation of the
one-loop contributions of chiral superfields in the diagrams
with external vector legs. It is known \cite{Gat} that such
contributions in the full EA are determined by the expression
${\rm Tr}\{\ln(\nabla^2\bar\nabla^2-m^2) + h.c.\} $
Using $\int d^4\theta=\int d^2\theta\bar\nabla^2,$ we obtain
the known basic chiral expression \be \int
d^2\theta\ln(\bar\nabla^2\nabla^2-m^2)\bar\nabla^2\delta^{(8)}+h.c.=
\int d^2\theta\ln(\Box_{+}-m^2)\bar\nabla^2\delta^{(8)},
\ee
where
$\Box_{+} =\Box-iW^\alpha\nabla_\alpha-i/2 (\nabla W)$ with covariant $\Box$.
The transition to the momentum representation consists in
replacement of the assumed operators and fields by
correspon\-ding pseudodifferential operators and the additional
integration $\int {d^4 p\over (2\pi)^4} d^2\psi d^2\bar\psi$.
Obviously, all $\partial_\psi,$ $\bar\partial_{\bar\psi}$
symbols from right-hand side of $\bar\nabla^2_p$ can be
omitted, since they act on nothing. Having in mind the property
of the Grassmanian integration, it is also possible to omit all
$\bar{\partial}_{\bar\psi} $ acting on
$\bar{\psi}^2$ inside the logarithm and to perform integration over $d^2\bar\psi$.

 %Inserted by TeXtelmExtel

Further, it is convenient to proceed to the proper-time
representation for the logarithm of operator and to use
the appropriate $\zeta$ regularization $\Gamma_{(1)} \sim-\zeta'
(0).$ The next step  in our strategy, which helps us to get the final
result practically without computations, consists in separation
of exponents of the operators $\nabla_p$ and the covariant 'box'
\be
K(T)= e^{-TD'}\int{d^4 p\over(2\pi)^4}d^2\psi  {\rm
e}^{-T\nabla^\alpha_p iW^p_\alpha} {\rm e}^{T\Box_{p} + ...},
\ee
where the omitted terms are $W\bar W$ and $W^2\bar W^2$,  since
the factor in front of the integral obviously will be $W^2$.
Moreover, in the considered $U(1)$ gauge effective theory, they
do not give the contribution. With the purpose to reduce the
problem of performing the trivial integration over $d^2\psi$, we
extract from the $T$ exponent 
$$ 
\exp T(\psi_\alpha
iW^\alpha+\psi_\alpha\partial^\beta N^{\alpha}_\beta), \quad
N^{\alpha}_{\beta}=if^{\alpha}_{\beta} + \delta^{\alpha}_{\beta}D',
$$
the operator of affine transformations, 
i.e., $\exp (\psi_{\alpha} \partial^{\beta})$. Using
$$
[\psi_\alpha\partial^\beta,\psi_\gamma]=\delta^\beta_\gamma
\psi_{\alpha},\quad \exp(\psi_\alpha \partial^\beta
N^{\alpha}_\beta)\cdot 1 = 1, $$ and the identity (\ref{Oper}),
we get 
$$ 
K(T)=\int{d^4 p\over(2\pi)^4}d^2\psi {\rm
e}^{-TD'}\exp \left\{ -iW^\alpha ({{{\rm e}^{TN}-1}\over
N})^\beta_\alpha\psi_\beta \right\} {\rm e}^{T\Box_{p}}= 
$$ 
\be = W^2
{\rm e}^{-TD'} {\rm tr}\left( {{\rm e}^{TN}-1 \over N} \right) \int{d^4
p\over(2\pi)^4}{\rm e}^{T\Box_{p}}.  
\ee 
The last factor is the
Schwinger result (\ref{Sch}) for a scalar loop. For calculations
of a factor which modify the Heisenberg - Euler Lagrangian, we
diagonalize the matrix $N$ and find directly
\be\label{sgam}
\Gamma_{(1)}= \int d^8z W^2 \int_{0}^{\infty}{dT \over T}{\rm e}^{-Tm^2}
{\cosh(TD') - \cosh(T {\cal H}_{-}) \over {D'}^2-{\cal H}_{-}^2}K(T)_{Schw},
\ee
where ${\cal H}_{-}$ was defined in eq. (\ref{ev}).
Note the coincidence of this result with the result of
\cite{Shiz,Art}, obtained by essentially different methods.
Certainly, there is an ultraviolet divergence, which can be
excluded with the help of a wave function renormalization. It is
important to note, that the corrections to $W^2$ contain
nonholomorphic, in the sense of Seiberg, terms $f_{\dot\alpha\dot\beta}.$
The superfield action (\ref{sgam}) reproduces correctly the
results of the calculations on the component level \cite{Vec}.

 %Inserted by TeXtelmExtel

In the last example we will consider contributions from only
the quantum gauge field $V$. After splitting the field into a
background and quantum part, the SYM action in Fermi - Feynman gauge is
$$
{\cal S}=-{1 \over 2g^2}{\rm Tr}[({\rm e}^{-V}\nabla^{\alpha}{\rm
e}^{V})\bar{\nabla}^{2}({\rm e}^{-V}\nabla_{\alpha}{\rm e}^{V})
+ V(\bar{\nabla}^{2}\nabla^{2} + \nabla^{2}\bar{\nabla}^{2})V ].
$$
 The quadratic action has the form
$$
{\cal S}_{0} = -{1 \over 2g^2}{\rm Tr}\left(V[\Box
-iW^{\alpha}\nabla_{\alpha} -
i\bar{W}^{\dot{\alpha}}\bar{\nabla}_{\dot{\alpha}} ]V\right).
$$
All the dependence on the background fields is through the
connection coefficients and through the background field
strength.  Further, we use the heat kernel representation of the
EA and change all quantities by pseudodifferential operators as
before.  In this case $\Box_V = \Box_p + i\nabla^p_\alpha
W^\alpha_p + i\bar\nabla^p_{\dot\alpha} \bar W^{\dot\alpha}_p.$
Here all one-loop background graphs are finite in super QCD
theories, but they potentially have an infrared singularity,
that is an attribute of an unstable mode.  We consider 
$U(1)$ gauge theory case. Following our strategy, we set all three
operators in separate exponents 
$$ 
{\rm e}^{T\nabla^p_\alpha
iW^\alpha_p} {\rm e}^{T\bar\nabla^p_{\dot\alpha} i\bar
W^{\dot\alpha}_p} L(W,\bar W){\rm e}^{T\Box_p}.  
$$ 
where
$L(W,\bar W)$ is the function of the superfields $W,\bar W$ and
the operator $\nabla^p_{\alpha\dot\alpha}.$ For SQED, where the
power $W$ is limited by 2, the function does not give the
contribution to the EA. Now, as well as in the previous example
of this section, we have nonzero contributions to $d^2\psi
d^2\bar\psi$ integrals
$$
\int d^2 \psi d^2\bar\psi
{\rm e}^{T\psi_\alpha iW^\alpha_p}
{\rm e}^{T\bar\psi_{\dot\alpha} i\bar W^{\dot\alpha}_p}
\int{d^4 p\over(2\pi)^4}{\rm e}^{T\Box_p}
$$
and we can, using results of the previous calculations,  
show at once the final result
\be\label{WW}
K(T)= W^2\bar{W}^2\det({{{\rm e}^{TN}-1}\over N})
\det({{{\rm e}^{T\bar{N}}-1}\over \bar{N}}){1\over(4\pi
T)^2}[\det{TF\over \sinh(TF)}]^{1/2},
\ee
where
$N^\beta_\alpha=iD_\alpha W^\beta,$ $\bar
 N^{\dot\beta}_{\dot\alpha}=i\bar D_{\dot\alpha}\bar
 W^{\dot\beta}.$ To check this result, we could use the
technique of correlator calculation \cite{Art}, which we have
already demonstrated in section 5.

 %Inserted by TeXtelmExtel

As well as for covariant constant YM background, the condition
$ [\nabla_{\alpha \dot{\alpha}}, W_{A}] = 0$ leads
to the anticommutator
$\{W_{\alpha}^{a},\bar{W}_{\dot{\alpha}}^{a}\}=0 $, i.e., in
this approximation the superfields $W,\bar{W} $ are effective
abelian, and we can use ÿthe results for the EA super QED with
certain changes.  Full DEEA on a SYM background and chiral
superfields both in adjoint and in fundamental representation
demands a more detailed consideration. The complication originates
from ${\cal S}^{2}_{mix} $ and mass terms in the operator
$\Box_V$, which depends on chiral fields.

 %Inserted by TeXtelmExtel

\newpage
\section{Summary} % 8

 %Inserted by TeXtelmExtel

 %Inserted by TeXtelmExtel

In the present paper we develop elegant and effective
technique based on noncommutative geometry of deformation
quantization for calculation of the expansion in the
derivatives of background fields for the one-loop effective action. 
It is important that the supersymmetrical and gauge invariant form is
conserved through all stages of calculations. We use the 
simple idea of exploiting
a canonical transformation that leads to the normal coordinate
expansion of symbols. It is the well known realization of
equivalence principle which requires the existence  of such a
reference frame at every point that the effects of gauge fields
can be locally neglected.

 %Inserted by TeXtelmExtel

To test the approach suggested  we focused
 on comprehensi\-vely investigated models, though all
constructions could be applied straightforwardly to the QFT
models involving difficulties in the quantization.  In all
examples considered, the results of the proposed computing
scheme coincide completely with the known ones.  The
suggested approach allows 'manual'
manipulations to be effectively replaced  
by computer methods to get all next HMDS
coefficients in the expansion of the one-loop effective action.

 %Inserted by TeXtelmExtel

It can be also said that the approach shows the problem from
another side and extends our knowledges about the structure of
the path integrals. Other applications of the presented method and its
modifications for nonflat and harmonic superspace will be given
in the subsequent papers.

 %Inserted by TeXtelmExtel

\section{Acknowledgments}
We thank V.G. Serbo and A.I. Vainstein for useful discussions.
The authors are also grateful to V.P. Gusynin and I.Avramidi for
informing us about several papers.  This work was partially
supported  by RFBR 96 - 02 - 19079 grant.

 %Inserted by TeXtelmExtel

\newpage

 %Inserted by TeXtelmExtel

 %Inserted by TeXtelmExtel

 %Inserted by TeXtelmExtel

\end{document}